# Ultrafast pulsed laser evaluation of Single Event Transients in opto-couplers

**Kavin Dave[1]\*, Aditya Mukherjee[1], Hari Shanker Gupta[2], Deepak Jain[1], and Shalabh Gupta[1]**

1. Department of Electrical Engineering, Indian Institute of Technology Bombay, Mumbai, India
2. Space Applications Center, Indian Space Research Organization, Ahmedabad, India

E-mail: kavindave22@gmail.com*

**Abstract:** We build a 1064 nm fiber laser system-based testing facility for emulating SETs in different electronics components and ICs. Using these facilities, we tested the 4N35 optocoupler to observe SETs for the first time.

## 1. Introduction

Optocouplers (also known as optoisolators) are used for switching and isolating high and low-voltage circuits. These features make them suitable for consumer electronics to space vehicles spaceflights, satellites, and planetary rovers. The basic build design of optocouplers involves a light source and a photosensitive detector placed in close proximity. This makes them susceptible to Galactic Cosmic Radiations (GCR) which can affect their operation and in turn operation of the device. The Single Event Transients (SETs) are one of the major radiation effects that need to be investigated for optocouplers. SETs cause temporary or permanent disruption of the normal functioning of an electronic component, induced by a single ionizing particle passing through a sensitive node on it. These temporary voltage fluctuations can induce logical errors in a digital circuit and failure of the device. Several such incidents have been reported in space applications, with two most famous being associated with Hubble Telescope and TOPEX/Poseidon spacecraft [1]. In order to emulate these radiation effects, several testing facilities have been built using particle accelerators. However, these facilities are sparse and require high operational and maintenance costs. NASA and different other organizations tested several different types of optocouplers including 4N35 using different accelerators at different facilities [2]. However, no SETs were reported for 4N35 optocouplers using particle accelerators [2].

The laser-based testing facilities to emulate these radiation effects have been reported, which are fairly easy to establish, operate, and maintain. We have also built such a facility at IIT Bombay in collaboration with the Indian Space Research Organization (ISRO) to test different electronic components and ICs. Yingqi *et* al. [3] have tested 4N49 and HCPL5231using a 20 ns pulsed laser. In this paper, for the first time, we report the laser testing of a 4N35 optocoupler using a 10 ps laser. It is pertinent to mention that the time scale of interaction between the ionizing particle and active layer of the semiconductor is on the scale of a few pico-second [4]. Therefore, a pico-second laser can offer better emulation of the SETs than a nano-second laser.

## 2. Experimental set-up

We built an in-house laser testing system at IIT Bombay using a commercially available 10 picosecond 1064 nm laser with a pulse repetition rate of 20 MHz. TH pigtail fiber of the laser has a lens on its face, ensuring the spot size of the light is 1 µm at the focus point. In this experiment, we tested commercially available 4N35 optocouplers employing GaAs infrared LED and an NPN Si phototransistor for photoconductance. Here the base-collector junction acts as the active region of the device. Figure 1(a) the experimental set-up and 1(b, c, d, e) shows the close-up pictures of the optocoupler. The base terminal is left unconnected ensuring that the base current consists of photocurrent. The ceramic capping of the optocoupler was dusted off until a translucent gel covering the die was observed, as shown in figure 1(b). While testing, LED was off to observe the laser-induced transients. Lastly, the experiments were performed in a dark room to avoid any contribution from the stray light. The voltage across the collector-emitter junction was recorded on a digital oscilloscope while varying laser power.

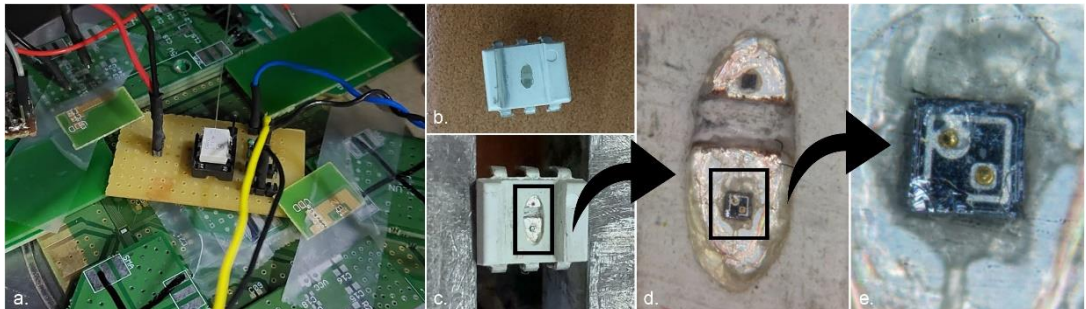

Figure 1 (a) Experimental set-up showing capped 4N35 and lensed fibre to focus laser spot (b) de-capped 4N35 with optical gel (c) optical gel removed (d) Internal structure of 4N35 containing LED and NPN Silicon phototransistor (e) Close-up of phototransistor

## 3. Results and Discussion

In absence of any light, there is only a thermal current flowing across the collector-emitter junction, called the dark current. When light falls on this collector-base junction, photons are absorbed in the depletion region, forming free electron-hole pairs which move under the influence of applied potential between collector and emitter. This forms the base current. This photo-generated base current amplifies the collector current.

The equivalent of linear energy transfer (LET) by laser beam depends on laser power and laser fluence. We calculate equivalent LET assuming that the electron-hole pairs generated in a sensitive volume of depth 'd' by ions and lasers are equal. The number of electron-hole pairs formed in sensitive volume by a laser of laser power $y$ watt is equal to the number of electron-hole generated by an ionizing particle beam of LET per unit mass density equal to $c$ (MeV.cm$^2$/mg).

$$\text{Eq LET of laser} = \frac{\text{e−h generated in active region by laser of y watt}}{\text{e−h pairs generated by of an ions of LET per unit mass denisty c ((MeV.cm\textasciicircum2/mg)}} \times c\ (\text{MeV.cm\textasciicircum2/mg})$$

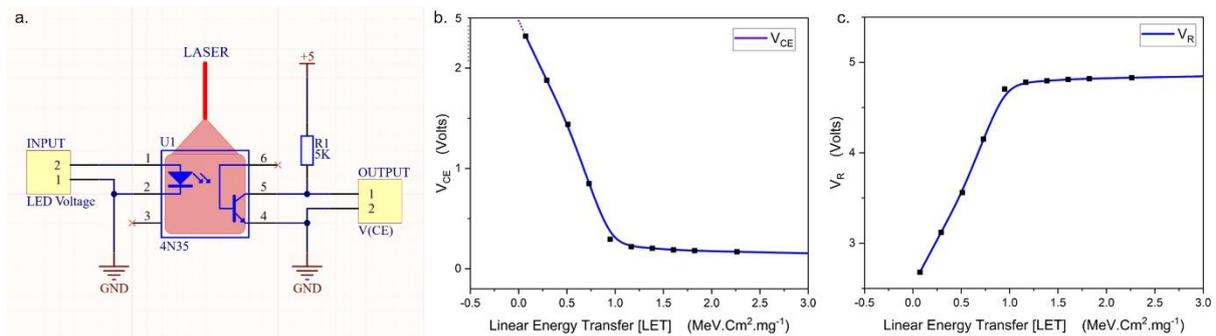

Figure 2 (a) Schematic of experimental set-up, (b) $V_{CE}$ vs. LET and (c) Voltage across R1 ($V_R$) vs. LET

For the 4N35 optocoupler, we observe that an ion of equivalent LET of 1.17 MeV.cm$^2$/mg drives the optocoupler into saturation. The threshold LET for which SETs can be observed was found to be $\leq 0.07$ MeV.cm$^2$/mg which is an extremely small value. In our experiment, we could not observe transient pulses for collector-emitter voltage. This is due to the much higher repetition of laser pulses (~20 MHz) than the rise and fall time of 4N35 optocouplers [5]. To observe transient pulse amplitude, pulse width, and SETs cross-section, the pulse repetition of the laser should be of the same order as that of the device response. However, from existing data on other optocouplers [3,6], it can be inferred that the pulse width of the transient pulse increases with an increase in equivalent LET until saturation of the device is reached. Maximum possible transient pulse width at saturation than would simply be equal to the time taken by generated electron-hole pairs to recombine, which is about 50 μs for 4N35 [5]. To obtain data on SETs cross-section, transient pulse width and pulse amplitude for slow optocouplers such as 4N35, a picosecond laser with a low repetition rate may be used or a laser chopper may be employed.

## 4. Conclusion

We have set up a 1064 nm laser-based testing facility for emulating SETs effects in electronics components and ICs. We have miniaturized the set-up by employing a lensed fiber patch cord ensuring the ~1 μm spot size at its focal point. To the best of our knowledge, this is the first report demonstrating the laser-based evaluation of SETs in the 4N35 optocoupler. We could calculate the threshold and saturating equivalent LET values for the 4N35 optocoupler.

## 5. References

[1] R.A. Reed *et al.*, "Emerging Optocoupler Issues with Energetic Particle-Induced Transients and Permanent Radiation Degradation" IEEE Trans Nucl Sci, Vol 45, No. 6, December 1998
[2] K.A. LaBel *et al.* "A compendium of recentoptocoupler radiation test data,". IEEE Radiation Effects DataWorkshop:123-146, July 2000.
[3] Ma Yingqi, "Pulsed Laser Evaluation of Single Event Transients in Optocouplers", Tencon 2009, Beijing, China
[4] S. P. Buchner *et al.* , "Pulsed Laser Testing for Single Event Effects Investigation", IEEE Trans Nucl Sci, Vol 60, No. 3, June 2013
[5] Datasheet, 4N35, Vishay Semiconductors, https://www.vishay.com/doc/?81181=
[6] A.H. Johnston *et al.*, "Single Event Upset Effects in Optocouplers ", IEEE Trans Nucl Sci, Vol 45, No. 6, December 1998.
[7] A.Javanainen *et al.*, "Linear Energy Transfer of Heavy Ions in Silicon", IEEE Trans Nucl Sci, Vol 54, No. 4, August 2007